\documentstyle[12pt]{article}
\begin{document}

\title{{\bf How to observe fluctuating temperature?}}
\author{O.V.Utyuzh$^{1}$\thanks{e-mail: utyuzh@fuw.edu.pl},
G.Wilk$^{1}$\thanks{e-mail: wilk@fuw.edu.pl} 
and Z.W\l odarczyk$^{2}$ \thanks{e-mail: wlod@pu.kielce.pl} \\
$^1${\it The Andrzej So\l tan Institute for Nuclear Studies}\\
{\it Ho\.za 69; 00-689 Warsaw, Poland}\\
$^2${\it Institute of Physics, Pedagogical University}\\
{\it Konopnickiej 15; 25-405 Kielce, Poland}}
\date{\today}
\maketitle

\begin{abstract}
We provide arguments that event-by-event (EBE) analysis of
multiparticle production data are ideal place to search for the
possible fluctuation of temperature characterizing hadronizing source
in thermodynamical approach.\\

PACS numbers: 25.75.-q 24.60.-k 05.20.-y 05.70.Ln
\end{abstract}

In this note we would like to advertise the feasibility and
importance of the observations of fluctuating temperature in high
energy multiparticle production processes. Let us first list our
chain of reasoning: 
\begin{itemize}
\item ALICE will look (among other things) for the creation of Quark
Gluon Plasma (QGP);
\item this calls for a thermodynamical description of (at least some)
relevant observables and this in turn calls for the temperature $T$ of
the system under investigation as one of the most important quantities;
\item $T$ can be deduced most directly (at least this is believed to
be the case) by looking at $p_T$ spectra because ($\mu_T = \sqrt{m^2 +
p_T^2}$)
\begin{equation}
\frac{dN}{dp_T}\, \propto \exp \left( - \frac{\mu_T}{T}
\right) . \label{eq:def}
\end{equation}
\end{itemize}

This is widely accepted approach, notwithstanding the fact that
both details of what $T$ really means or whether (\ref{eq:def}) is a
proper form for $\mu_T$-dependence are still subject to hot debate
and modelling. Taking therefore (\ref{eq:def}) as our starting point
we want to concentrate on question: is it possible that $T$ is
fluctuating quantity \cite{GEN,QCD} and if so, does it fluctuates
only from event-to-event or also in a given event? \\

We are not going to discuss here the problem of internal consistency
(or inconsistency) of the notion of fluctuations of temperature in
thermodynamics referring in this matter to \cite{FLUCT,L,OTHER}.
What we want to do is to brings to ones attention the fact that
event-by-event analysis allows (at least {\it in principle}) to
detect fluctuations of temperature taking place {\it in a given
event}. This is more than indirect measure of fluctuations of $T$
proposed some time ago in \cite{KM} or more direct fluctuations of
$T$ {\it from event to event} discussed in \cite{GEN}.\\

To this end let us first remind our results presented in \cite{WW}
where we have shown that fluctuation of the parameter in exponential
distribution leads in a  natural way (under some circumstances, of
course, to be mentioned in a moment) to final distribution of the
power-like form (know also as L\'evy distribution): 
\begin{equation}
\left\langle \exp\left[ - \left(\frac{1}{T}\right)\cdot \mu_T\right]
\right\rangle \, \Longrightarrow \, \left[ 1 +
(1-q)\left(\frac{1}{T_0}\right) \cdot \mu_T\right]^{\frac{1}{1-q}} .
\label{eq:fluct} 
\end{equation}
Here averaging $\langle \dots\rangle$ is performed over fluctuations
of parameter (here our temperature) $\frac{1}{T}$ which take place
around some mean value $T_0$ and should follow gamma distribution 
(see \cite{WW} for details). The new parameter occuring here
(identical to the so called entropic index or nonextensivity
parameter in Tsallis statistics \cite{T}) is tightly connected with
the size of such fluctuations, namely
\begin{equation}
q\, =\, 1\, +\, \omega \label{eq:q}
\end{equation}
where 
\begin{equation}
\omega\, =\, \frac{\left\langle\left(\frac{1}{T}\right)^2\right\rangle
                 - \left\langle \frac{1}{T}\right\rangle^2}{\left\langle \frac{1}{T}\right\rangle^2}
.\label{eq:omega} 
\end{equation}

It is worth to mention that distribution of the L\'evy type
(\ref{eq:fluct}) has been observed already in inclusive processes
\cite{ALQ}. However, inclusive processes are not able to provide
unambiguous answer what is the source of such behaviour. This can be
done, such is our belief, only in the careful analysis of
event-by-event data, especially those for heavy ion collisions.  
Two scenarios are possible here and should be subjected to 
experimental verification:
\begin{itemize}
\item[{\bf (1)}] $T$ is constant in each event but because of
different initial conditions it fluctuates from event to event.
In this case in each event one should find exponential dependence
(\ref{eq:def}) with $T=T_{event}$ and possible departure from it will
occur only after averaging over all events. It will reflect
fluctuations originating in different initial conditions for each
collision from which given event originates. This situation is
illustrated in Fig. 1 where $p_T$ distributions for $ T = 200$ MeV
(black symbols) and $T = 250$ MeV (open symbols) are presented. All
other details are the same as listed below for Fig. 2. Such values of
$T$ correspond to typical uncertainties in $T$ expected at LHC due to
different initial conditions. Notice that both curves presented here
are straight lines.
\item[{\bf (2)}] $T$ fluctuates in each event around some value
$T_0$. In this case one should observe departure from the exponential
behaviour already on the single event level which should be fully
given (\ref{eq:fluct}) with $q>1$. It reflects situation when, due
to some intrinsically dynamical reasons, different parts of a given
event can have different temperatures \cite{WW}. Fig. 2 shows
typical event of this type obtained in simulations performed for
central $Pb+Pb$ collisions taking place for beam energy equal 
$E_{beam}=3~A\cdot$TeV in which density of particles in central 
region (defined by rapidity window $-1.5 <y< 1.5$) is equal to 
$\frac{dN}{dy} = 6000$ (this is the usual value given by event 
generators like VENUS, SHAKER, HIJING). Black symbols represent 
exponential dependence obtained for $T = 200$ MeV (the same as 
in Fig. 1), open symbols show the power-like dependence as given 
by (\ref{eq:fluct}) with the same $T$ and with $q=1.05$ (notice 
that the corresponding curve bends slightly upward here). In this 
typical event we have $\sim 18000$ secondaries, i.e., practically 
the maximal possible number. Notice that points with highest $p_T$ 
correspond already to single particles.
\end{itemize}

One should stress here the following important fact: our $\omega = q
- 1$ has physical meaning of the total heat capacity $C$, because
according to a  basic relation of thermodynamic \cite{L} ($\beta =
\frac{1}{T}$)  
\begin{equation}
\frac{\sigma^2(\beta)}{\langle \beta\rangle ^2}\, =\, \frac{1}{C}\, 
=\, \omega\, =\, q - 1 .\label{eq:C}
\end{equation}
Therefore measuring in addition to the temperature $T$ also 
nonextensivity $q$ describing its fluctuation (and, because of this,
the total heat capacity $C$) could be of great practical importance
for our understanding of dynamics of heavy ion collisions
\cite{GEN,QCD}. In particular it should not only facilitate checking
the commonly made assumption that an approximate thermodynamics state
is obtained in a single collision but also, by knowing the heat
capacity, provide considerable information about its thermodynamics
(especially on the existence and type of the possible phase
transitions \cite{GEN,QCD}) \cite{FOOT}.\\

This work was partially supported by Polish Committee for Scientific 
Research (grants 2P03B 011 18 and  621/E-78/SPUB/CERN/P-03/DZ4/99).\\

\newpage
\begin{figure}[t]
\setlength{\unitlength}{1cm}
\begin{picture}(23,16)
\includegraphics{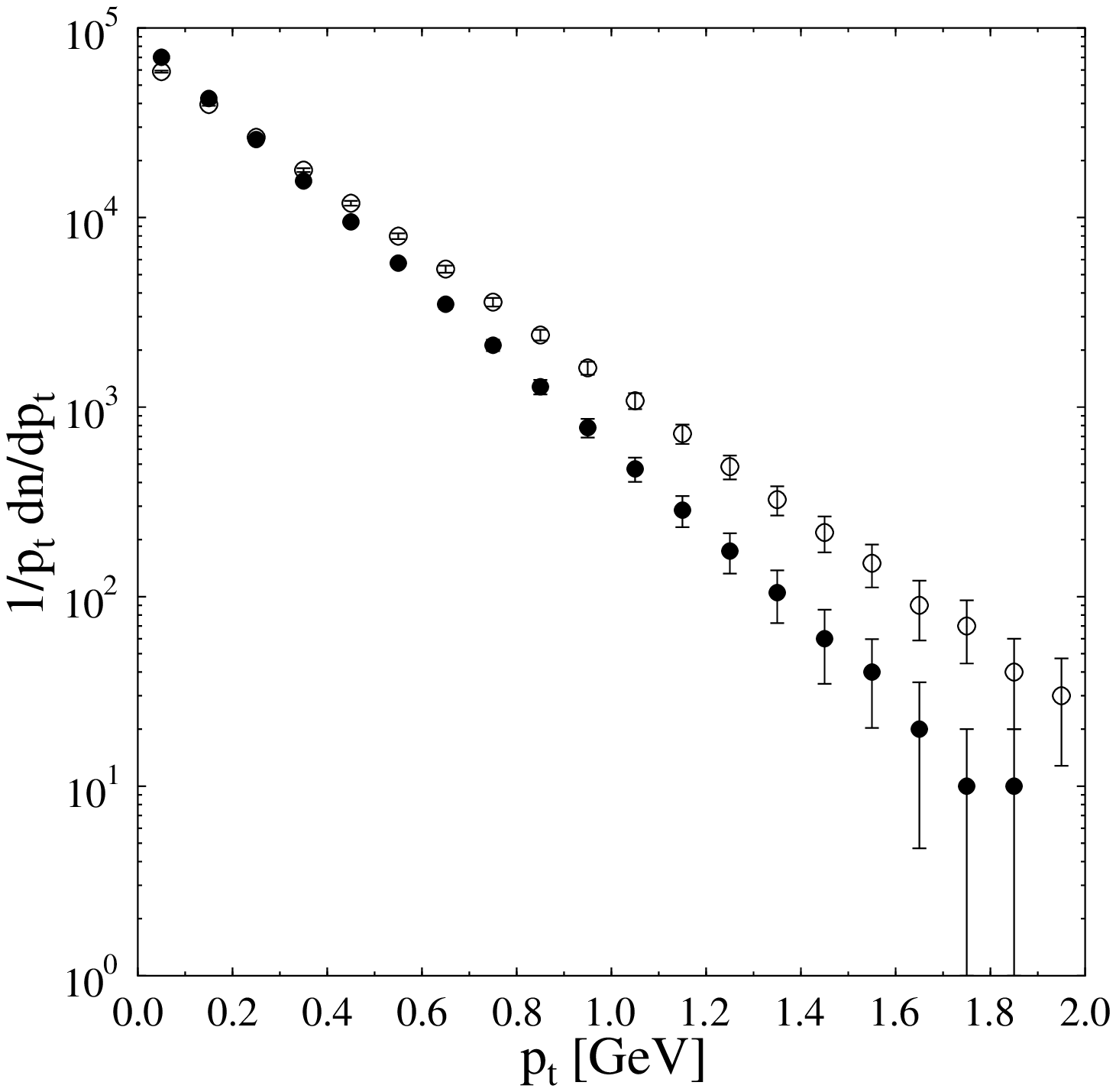}
\end{picture}
\end{figure}
\centerline{Figure 1}

\newpage
\begin{figure}[t]
\setlength{\unitlength}{1cm}
\begin{picture}(25.,16.5)
\includegraphics{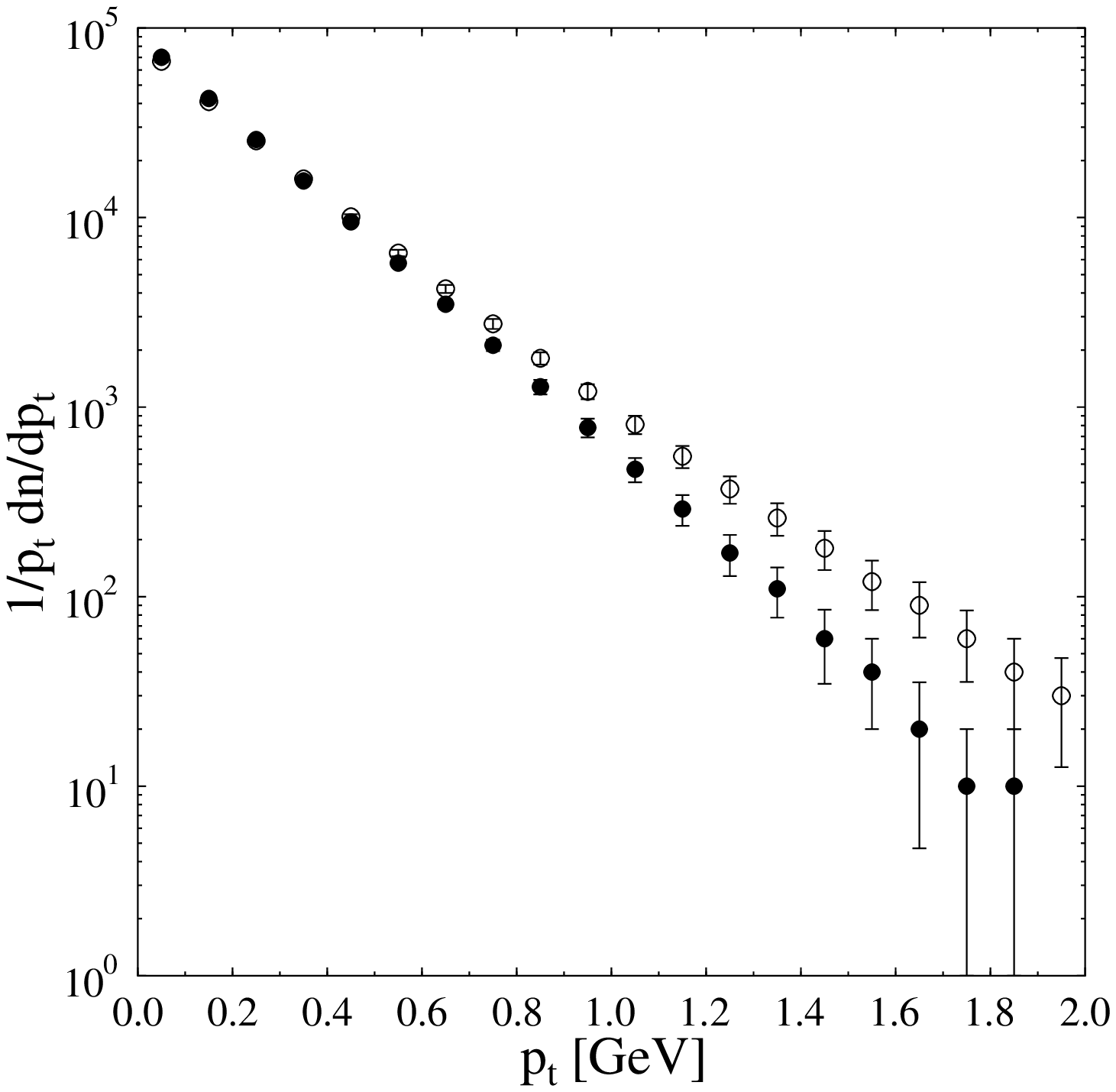}
\end{picture}
\end{figure}
\centerline{Figure 2}

\begin{thebibliography}{99}
\bibitem{FOOT} It is interesting to realize that for the Planckian
                gas at $T=186$ MeV, occupying volume of the order of
                the volume of sulfur nucleus, one gets $C=34.4$ per 
		degree of freedom, which leads, using (\ref{eq:C}), 
		to $q=1.015$ obtained for such system for the $p_T$ 
		dependence of produced secondaries.
\end{thebibliography}
\end{document}